# A background simulation method for cosmogenic nuclides inside HPGe detectors for rare event experiments

Jian Su, Zhi Zeng, Hao Ma, Qian Yue[1], Jianping Cheng, Jin Li

Key Laboratory of Particle and Radiation Imaging (Ministry of Education) and Department of Engineering Physics, Tsinghua University, Beijing 100084

ABSTRACT: Cosmogenic nuclides inside germanium detectors contribute background noise spectra quite different from ordinary external sources. We propose and discuss a nuclide decay and level transition model based on graph theory to understand the background contribution of the decay of cosmogenic nuclides inside a germanium crystal. In this work, not only was the level transition process, but the detector response time was also taken into consideration to decide whether or not to apply coincidence summing-up. We simulated the background spectrum of the internal cosmogenic nuclides in a germanium detector, and found some unique phenomena caused by the coincidence summing-up effect in the simulated spectrum. Thus, the background spectrum of each cosmogenic nuclide can be quantitatively obtained.

## 1 Introduction

With the advantages of very low radioactivity, high energy resolution, and low energy threshold, High Purity Germanium (HPGe) is used as the detector material for CDEX[1, 2], CoGeNT [3], GERDA [4], and other experiments which search for dark matter or neutrinoless double-beta (0νββ) decays. All these experiments run in underground laboratories, in order to reduce the background caused by cosmic rays. During the process of growing and fabricating an HPGe crystal, and its manufacture into a germanium detector, all of which happens at ground level, the HPGe crystal will be exposed to cosmic rays, resulting in some cosmogenic radioactive nuclides being produced inside the crystal. Although the short-lived radioactive nuclides decrease dramatically due to their short half-lives, the long-lived radionuclides will contribute to the background continuously. These long-lived radionuclides inside the germanium crystals include $^{68}$Ge ($T_{1/2}$=288d), $^{65}$Zn ($T_{1/2}$=244d), $^{60}$Co ($T_{1/2}$=5.27y), $^{58}$Co ($T_{1/2}$=71d), $^{57}$Co ($T_{1/2}$=270d), $^{56}$Co ($T_{1/2}$=77.3d), $^{63}$Ni ($T_{1/2}$=100.1y), $^{55}$Fe ($T_{1/2}$=2.73y), $^{54}$Mn ($T_{1/2}$=312d), and so on [5]. All these cosmogenic long-lived radionuclides can contribute to the background in both the low and high energy regions.

In previous works, the X-ray peaks of cosmogenic radionuclides inside the germanium detector between low-threshold (~several keV) to 12 keV have been observed and evaluated by the CDEX[6], CoGeNT [7] and CDMS-II [8] experiments. So it is possible to evaluate the background contribution for these cosmogenic nuclides to the whole energy region, especially the high energy region in which the 0νββ experiments are interested.

The internal cosmogenic nuclides contribute to the background with their own distinct features. Generally, more than one particle, such as X-rays, β$^+$, β$^−$, and/or γ, will be emitted in the decay process of cosmogenic nuclides. If more than one particle is generated within the detector response time, their total energy deposition will be collected as one event, which is called the "summing-up effect" in gamma spectrometry. However, the time between the particles being emitted in a decay process is quite short (~ps or ~ns), so the energy deposition always tends to sum up. A spectrum with the summing-up effect is quite different from the spectrum of independently emitted particles without the summing-up effect. To solve this problem, we modeled the CDEX-1 germanium detector with Geant4 code and simulated the background contributed by these internal cosmogenic nuclides in both the low and high energy regions. We also found some unique new features, such as the "X-ray barrier", caused by a coincidence summing-up effect in the simulated spectrum.

---
[1] Corresponding author. E-mail: yueq@mail.tsinghua.edu.cn

In this paper, we describe and discuss a method to simulate the background in the high energy region, based on the spectrum we have already measured from the low energy region, dealing especially with the cascade particle source in the decay and level transition process. In Section 2, we summarize the particle cascade mode of the cosmogenic nuclides inside a germanium crystal and the decay scheme data available. In Section 3, we explain how to solve the branching probability and the simulation principle of the cascade particles in detail. In the following sections, we introduce the unique phenomenon in the simulated spectrum caused by the coincidence summing-up effect inside a germanium crystal, and the feature of the contribution of the cosmogenic nuclides.

## 2 Simulation method for cosmogenic nuclide decays

### 2.1 Cosmogenic radionuclides and their decay schemes

The characteristic X-rays of the cosmogenic nuclides such as $^{73}$As, $^{68,71}$Ge, $^{68}$Ga, $^{65}$Zn, $^{56}$Ni, $^{56,57,58}$Co, $^{55}$Fe, $^{54}$Mn, $^{51}$Cr, and $^{49}$V have been clearly observed with the germanium detectors in the CoGeNT and CDEX experiments. As an important cosmogenic background source in $0\nu\beta\beta$ searches, $^{60}$Co is also taken into account, though there is no X-ray peak in the low energy region of less than 12 keV.

Over the past few years, decay-data evaluators have worked in an informal collaboration to create well-documented, high-quality evaluations of nuclear decay data [9]. We mainly use this decay scheme data to calculate level transition branching probability. We also use electron binding energy data from the X-ray data booklet [10]. Using this decay scheme, we list the nuclides with their possible cascade modes in Table 1. In the table, "X only" indicates that the nuclide may undergo electron conversion (EC) decay and directly arrive at the ground state of its daughter, so no γ will be emitted. "X + γ" indicates that the nuclide may undergo EC decay and arrive at an excited state of its daughter, and one or several γ-rays will follow. Similarly, "β$^+$ only", "β$^+$ + γ" and "β$^-$ + γ" tell us whether there are γ emissions following positron or electron emission.

**Table 1** Decay types of cosmogenic nuclides inside a germanium crystal

| Nuclides | Decay type | Possible cascade modes | | | | |
|---|---|---|---|---|---|---|
| | | X only | X + γ | β$^+$ only | β$^+$ + γ | β$^-$ + γ |
| $^{73}$As | EC | | ✓ | | | |
| $^{71}$Ge | EC | ✓ | | | | |
| $^{68}$Ge | EC | ✓ | | | | |
| $^{68}$Ga | EC or β$^+$ | ✓ | ✓ | ✓ | ✓ | |
| $^{65}$Zn | EC or β$^+$ | ✓ | ✓ | ✓ | | |
| $^{56}$Ni | EC or β$^+$ | ✓ | ✓ | ✓ | ✓ | |
| $^{56}$Co | EC or β$^+$ | | ✓ | ✓ | ✓ | |
| $^{57}$Co | EC | | ✓ | | | |
| $^{58}$Co | EC or β$^+$ | ✓ | ✓ | ✓ | ✓ | |
| $^{60}$Co | β$^-$ | | | | | ✓ |
| $^{55}$Fe | EC | ✓ | | | | |
| $^{54}$Mn | EC | ✓ | ✓ | | | |
| $^{51}$Cr | EC | ✓ | ✓ | | | |
| $^{49}$V | EC | ✓ | | | | |

From Table 1, we can find the important principle that there is no "pure γ" source inside germanium. The γ must followed an X-ray, a β$^+$-ray or a β$^-$-ray, and this is the main reason why the spectrum produced by some internal cosmogenic nuclide sources is quite different from that of the ordinary external source. This will be explained in more detail in Section 2.4.

## 2.2 "Probability tree" of energy level transitions

How many possible paths are there in the decay and level transition process for a radioactive nuclide? In this work, we use some concepts of graph theory [11] to describe our decay process model and algorithm. Taking $^{60}$Co as an example, its β$^-$ decay and the following level transitions are regarded as a five-level process. In the language of graph theory, the daughter nuclide's energy levels (levels 0,1,2,3) and the parent nuclide itself (equivalent to level 4) are regarded as vertexes and the decay or transitions linking the levels as edges in a decay scheme, as shown in Fig. 1.

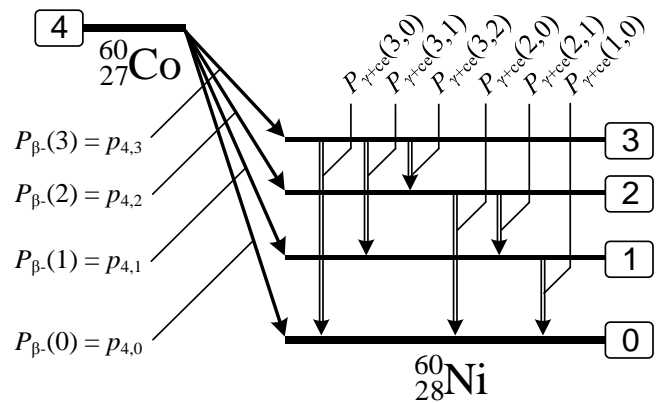

**Fig. 1.** Decay scheme of $^{60}$Co, showed the emission probabilities

We change this decay scheme to a probability tree (Fig. 2), called *Tree*(4), which contains all possible paths from level 4 to level 0.

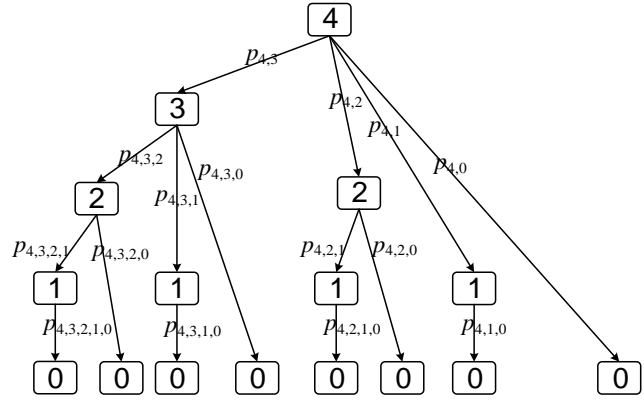

**Fig. 2.** Probability tree of a five-level decay, i.e. *Tree*(4)

In the probability tree, each arrow connecting two levels represents a transition. The parameter "$p$" beside each arrow represents the probability of that transition. The probability entering a vertex is equal to the sum of the probabilities leaving the vertex, e.g. for the vertex at level 3, $p_{4,3} = p_{4,3,0} + p_{4,3,1} + p_{4,3,2}$. The probability entering level 0 is the probability of finishing the transition path, called the "path probability".

Each whole decay and transition process must start from the highest level 4 and end at level 0, which is regarded as a path. There are 8 vertexes at level 0 in Fig. 2, indicating 8 paths in total.

The emission probability of each γ-ray (with its conversion electron) is given in the decay scheme, i.e. $P_{\gamma+ce}$. For example, there are two paths containing the edge 2→0 in Fig. 2, so the sum of their path probabilities is equal to the emission probability, i.e. $P_{\gamma+ce}(2,0) = p_{4,3,2,0} + p_{4,2,0}$. We can then establish the relationship between the emission probabilities and path probabilities as follows:

$$\begin{cases} P_{\gamma+ce}(3,0) = p_{4,3,0} \\ P_{\gamma+ce}(3,1) = p_{4,3,1} \\ P_{\gamma+ce}(3,2) = p_{4,3,2} \\ P_{\gamma+ce}(2,0) = p_{4,3,2,0} + p_{4,2,0} \\ P_{\gamma+ce}(2,1) = p_{4,3,2,1} + p_{4,2,1} \\ P_{\gamma+ce}(1,0) = p_{4,3,2,1,0} + p_{4,3,1,0} + p_{4,2,1,0} + p_{4,1,0} \end{cases} \quad (1)$$

From basic nuclear physics, in a transition process, a nucleus won't "remember" which energy level it comes from, so the transition steps are independent of each other. Each path probability is the product of its step probabilities as follows:

$$\begin{cases} p_{4,3,0} = p_{4,3} \times p_{3,0} \\ p_{4,3,1} = p_{4,3} \times p_{3,1} \\ p_{4,3,2} = p_{4,3} \times p_{3,2} \\ p_{4,3,2,0} = p_{4,3} \times p_{3,2} \times p_{2,0} \\ p_{4,2,0} = p_{4,2} \times p_{2,0} \\ p_{4,3,2,1} = p_{4,3} \times p_{3,2} \times p_{2,1} \\ p_{4,2,1} = p_{4,2} \times p_{2,1} \\ p_{4,3,2,1,0} = p_{4,3} \times p_{3,2} \times p_{2,1} \times p_{1,0} \\ p_{4,3,1,0} = p_{4,3} \times p_{3,1} \times p_{1,0} \\ p_{4,2,1,0} = p_{4,2} \times p_{2,1} \times p_{1,0} \\ p_{4,1,0} = p_{4,1} \times p_{1,0} \end{cases} \qquad (2)$$

Thus, the relationship between decay scheme data and transition step probabilities has been established. Note that each $p_{4,*}$ is given in the decay scheme, and is equal to the emission probability of each β$^-$-ray, i.e. $P_{\beta^-}(*)$ in Fig. 1.

### 2.3 Transition step probability calculation

Substituting equation ( 2 ) into equation ( 1 ), we can then get with a set of 6 equations with 6 unknown parameters, which are $p_{3,0}$, $p_{3,1}$, $p_{3,2}$, $p_{2,0}$, $p_{2,1}$, and $p_{1,0}$. Generally, if the highest level is $N$, the number of equations depends on the number of γ-rays, which is $\binom{N}{2} = \dfrac{N!}{2!(N-2)!}$. Since the number of unknown step probabilities is the same as the number of equations, there must be a unique solution.

However, the equation group is nonlinear, and so in this section we suggest an algorithm to solve it. Based on a single transition process, if γ($i,j$) is emitted, the nucleus should arrive at level $i$ and then transit directly to level $j$. First, we calculate the probability of the nucleus arriving at level $i$. Suppose the highest level is $N$. We use $P(i)$ to denote the probability of all possible routes from level $N$ to level $i$ (including "$N \to i$" and all kinds of "$N \to \ldots \to i$"). Note that $p_{i,j}$ denotes the probability of a transition from level $i$ to level $j$ directly, i.e. "$i \to j$". Thus we can get

$$P_{\gamma+ce}(i,j) = P(i) \cdot p_{i,j} \qquad (3)$$

The levels higher than $i$ are level $N$, ($N$-1), ... , and ($i$+1). From each of these, the nucleus can transit to level $i$. So $P(i)$ can be written as

$$P(i) = P(N) \cdot p_{N,i} + P(N-1) \cdot p_{N-1,i} + \ldots + P(i+1) \cdot p_{i+1,i} \qquad (4)$$

Equation ( 4 ) is actually a recurrence formula. Note that $p_{N,*}$ is known from the decay scheme, and $P(N) \equiv 1$. Using equation ( 4 ) and ( 3 ) in turn, we can solve $P(N$-$1)$, $P(N$-$2)$, ... , $P(1)$ one by one, with all step probabilities $p_{(N-1),*}$, $p_{(N-2),*}$, ... , $p_{1,*}$. The recurrence process is

$$
\begin{aligned}
&\text{for } i = (N-1) \text{ downto } 1 \\
&\{ \\
&\qquad P(i) = \sum_{k=i+1}^{N} P(k) \cdot p_{k,i} \\
&\qquad \text{for } j = 0 \text{ to } (i-1) \\
&\qquad\qquad p_{i,j} = \frac{P_{\gamma+ce}(i,j)}{P(i)} \\
&\}
\end{aligned}
\qquad (5)
$$

## 2.4 Source particle generation and transportation

Cosmogenic nuclides inside a germanium crystal may emit five kinds of particle when they decay: X-ray (low energy photon), $\beta^+$-ray (continuous energy positron), $\beta^-$-ray (continuous energy electron), $\gamma$-ray (high energy photon), and conversion electron (single energy electron) (see Table 1).

### 2.4.1 X-rays

For an ordinary external source, the X-ray peaks of the measured spectrum are K-shell or L-shell emission lines. An internal source, however, is different. The energy of the internal X-ray peaks is electron binding energy, which is a little higher than the energy for K-shell or L-shell emission lines.

The reason is that when EC decay occurs, an orbital electron is captured by the atomic nucleus, resulting in a hole. An outer electron will immediately transit to fill the hole and emit an X-ray or Auger electron, but will also leave a new hole. Another outer electron will then fill the hole and emit another X-ray. The whole process may result in the emission of several X-rays and/or Auger electrons. The whole series of these X-rays and Auger electrons may be quite complex, but fortunately their total energy is equal to the binding energy of the captured electron.

In addition, because the energies of the X-ray photons and Auger electrons are quite low (<10 keV), it is nearly impossible for them to escape from germanium crystal, so they always deposit all of their energy in the germanium detector and form a peak (except for the rare situation where they are emitted very close to the surface of the crystal). So there is no need to generate the whole series of X-rays and Auger electrons – we can just simulate a single photon whose energy is equal to the electron binding energy. In other words, the EC decay in our simulation generates only one photon, but the photon's energy is sampled from the K-, M-, or L-shell electron binding energy according to the probabilities based on the decay scheme data, and the energy is unrelated to which level it decays to. For example, an EC decay of $^{68}$Ga will emit an X-ray to reach $^{68}$Zn. The EC decay scheme and the data of X-rays are shown in Fig. 3 and Table 2, respectively.

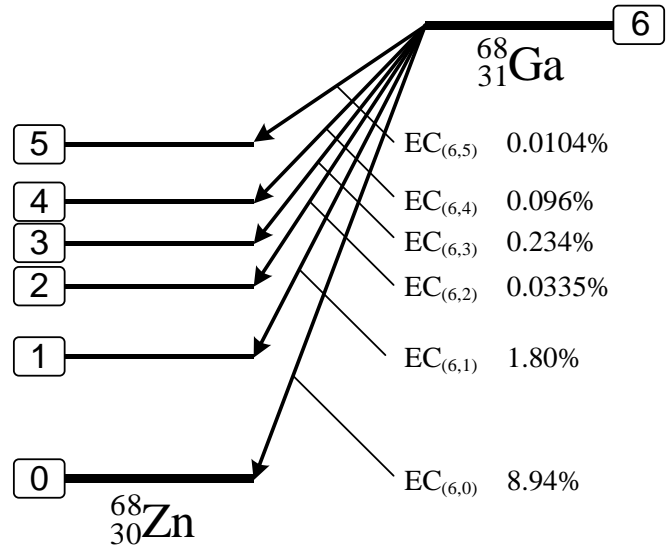

**Fig. 3.** EC decay scheme of $^{68}$Ga

**Table 2** X-ray peak energy (electron binding energy) of $^{68}$Zn and branching probabilities

|  | Branching probability | | |
| --- | --- | --- | --- |
|  | K-capture (9.659keV) | L-capture (1.1962keV) | M-capture (0.1398keV) |
| $EC_{(6,5)}$ | 0.8653 | 0.1141 | 0.0192 |
| $EC_{(6,4)}$ | 0.8823 | 0.1000 | 0.0166 |
| $EC_{(6,3)}$ | 0.8836 | 0.0989 | 0.0164 |
| $EC_{(6,2)}$ | 0.8839 | 0.0986 | 0.0163 |
| $EC_{(6,1)}$ | 0.8844 | 0.0983 | 0.0162 |
| $EC_{(6,0)}$ | 0.8847 | 0.0980 | 0.0162 |

2.4.2  Positrons and electrons

β decay emits a positron or an electron with a continuous energy distribution. The maximal energy of positrons or electrons can be found from the decay scheme, but the spectrum of each β decay is not given. For example, an β$^+$ decay of $^{68}$Ga will emit an positron to reach $^{68}$Zn, as the β$^+$ decay scheme shown in Fig. 4.

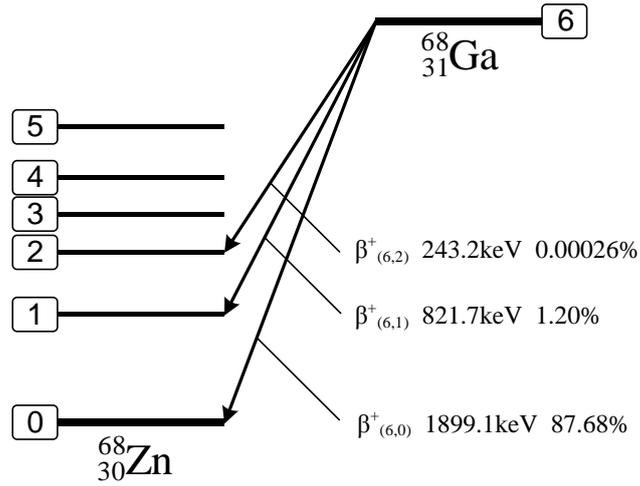

**Fig. 4.** β+ decay scheme of $^{68}$Ga

In this work, we use equation ( 6 ) [12] to evaluate the probability density function for the kinetic energy of the positron or the electron from each β+ / β- decay.

$$f(E_k) \propto \sqrt{E_k^2 + 2E_k m_e c^2}(E_{k\max} - E_k)^2(E_k + m_e c^2)F(Z', E_k) \qquad (6)$$

where

$$m_e = 0.511 \times 10^6,$$
$$c = 1,$$
$$F(Z', E_k) = \frac{x}{1 - e^{-x}},$$
$$x = \mp \frac{2\pi Z' c}{137 v} \text{ ( for } \beta\pm \text{ decay )},$$

$Z'$ is the atomic number of the daughter nucleus,

$$v = \frac{\sqrt{E_k^2 + 2E_k m_e c^2}}{E_k + m_e c^2}.$$

### 2.4.3 Gamma rays and conversion electron

A level transition will emit a γ-ray or conversion electron (CE). Generally, γ-rays have some probability of producing a CE instead. The probability is given by the decay scheme. A conversion electron is mono-energetic, and its energy is equal to the gamma energy minus the electron binding energy. After CE emission, X-rays will also be emitted. As in Section 2.4.1, we can regard the X-ray energy as equal to the electron binding energy.

### 2.5 Level transition time sampling

Each excited state of a nucleus has a half-life, so there are time intervals between the particles of a decay and transition process. The time interval can be regarded as the level transition time, which has an exponential distribution characterized by a single parameter λ. The relationship between λ and $T_{1/2}$ is

$$\lambda = \frac{\ln 2}{T_{1/2}} \qquad (7)$$

In the simulation process for decay and transition, each time the nucleus arrive at a level, the transition time should be sampled.

## 2.6 Detector response time and coincidence summing-up criterion

If more than one particle is generated within the detector response time, the total energy deposition from all the particles is added into the same event; this is called "coincidence summing-up". So in this work, the transition time is accumulated step by step, from the first particle of each decay process until the nucleus arrives at level 0 or until the cumulative time exceeds the detector response time. If the latter situation occurs, the energy deposition is recorded to the spectrum and initialized. The remaining processes of the unterminated transition are regarded as a new event.

Except for $^{73}$As, the other cosmogenic nuclides inside a germanium crystal have quite short half-lives for each level (of the order of fs, ps or ns), so their whole transition process is always shorter than the detector response time, which is about 30 μs for the CDEX point-contact germanium (PCGe) detector. This results in the particles of each decay always being likely to sum-up. For $^{73}$As, however, the half-life of its level 2, which is 0.499 s, is much longer than the detector response time. Its X-ray and γ-ray events, therefore, will not necessarily sum-up. Thus, the detector response time will significantly affect the spectrum shape of $^{73}$As.

# 3 Results and analysis

## 3.1 Features of spectrum with X-ray and γ coincidence summing-up (e.g. $^{54}$Mn)

Based on the method above, the background spectrum of internal $^{54}$Mn has been simulated, shown in Fig. 5. Three kinds of X-rays and only one kind of γ-ray are emitted in the decay of $^{54}$Mn, but six peaks can be found in Fig. 5(a). Three of the peaks are in the X-ray energy region, the other three are in the higher gamma energy region. Fig. 5(b) shows the low energy region from (a), with the three X-ray peaks: K-shell, L-shell, M-shell. The hatched portion 1 is a part of the Compton platform. This doesn't start from zero energy because the K-shell X-ray sums-up with the γ-ray. Regardless of whether the γ photon deposits energy in the detector or not, the K-shell X-ray will first deposit 5.989 keV, shifting the whole Compton platform to the right. Similarly, hatched portions 2 and 3 are parts of the Compton platform which sum-up with the L-shell and M-shell X-rays, respectively. Fig. 5(c) shows the high energy gamma peaks from (a). For the same reason as (b), the gamma peak splits into three parts, resulting in three "X + γ" sum peaks. The grey curves in Fig. 5(a), (b) and (c) are the spectra adding the energy resolution obtained from the CDEX-1 experiment[6].

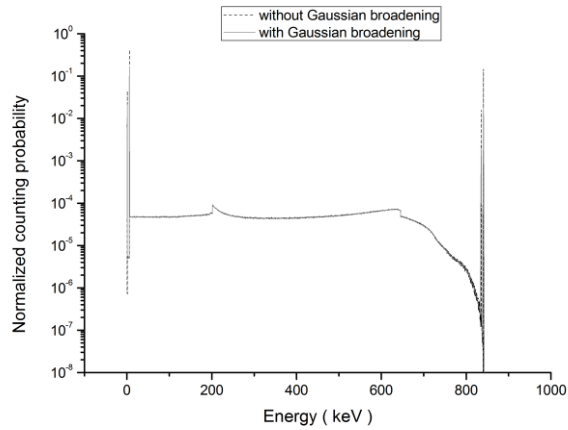

(a)

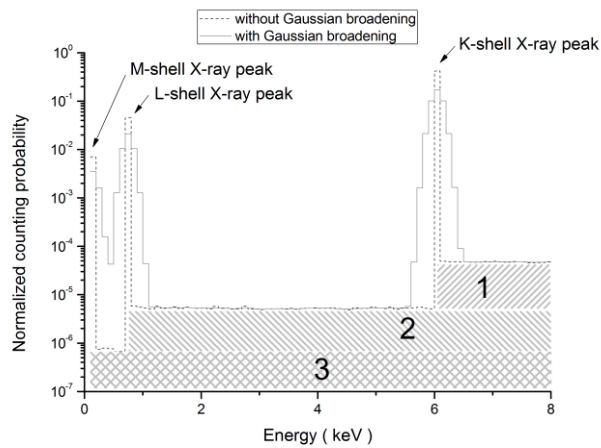

(b)

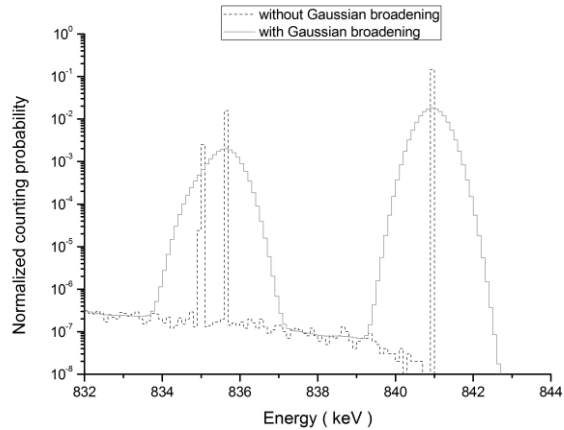

(c)

**Fig. 5.** Simulated spectrum of $^{54}$Mn

3.2 Features of spectrum with β⁻ and γ coincidence summing-up (e.g. $^{60}$Co)

Fig. 6(a) is the simulated background spectrum of internal $^{60}$Co with both β⁻ and γ cascade source. We can find that the spectrum is somehow similar to both γ spectrum and β⁻ spectrum, especially in the energy region of the full-energy peak and the sum peak. Fig. 6(b) is the simulated spectrum of $^{60}$Co with only the β⁻ source. Fig. 6(c) is the simulated spectrum of $^{60}$Co with only the γ cascade source. We can see two full-energy peaks at 1173 keV and 1332 keV. Their sum peak is also clear. The single and double escape peaks for the first and second full-energy peaks, as well as for the sum peak, are also seen. Finally, the annihilation peak at 511 keV is seen.

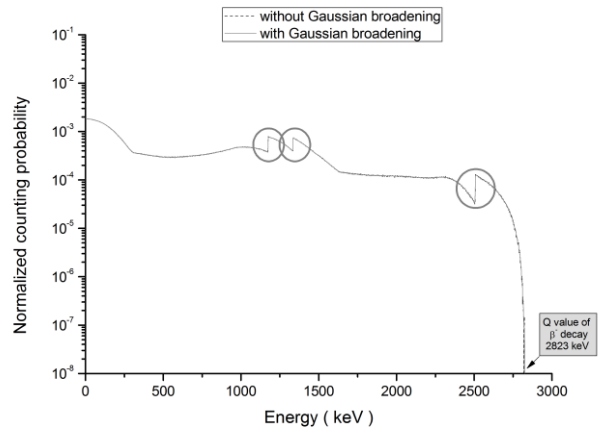

(a)

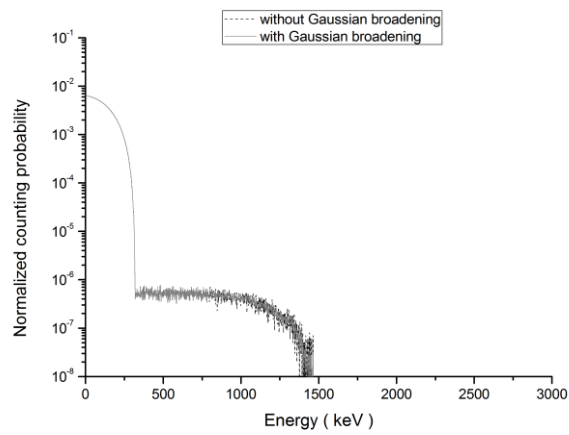

(b)

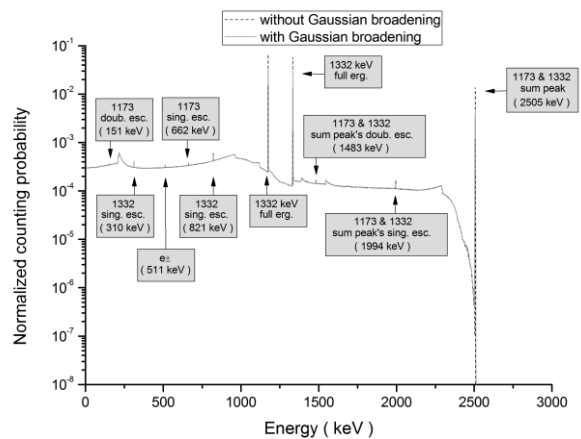

(c)

**Fig. 6.** Simulated spectrum of $^{60}$Co

## 3.3 Simulated spectrum of other cosmogenic nuclides (e.g. $^{73}$As and $^{68}$Ge)

Fig. 7 is the simulated background spectrum of internal $^{73}$As with both X-ray and γ cascade source. Three kinds of X-rays and two kinds of γ-rays are emitted in the decay of $^{73}$As, but we can see seven peaks in Fig. 7 (sum peaks associated with M-shell X-ray actually exist but aren't discussed here for their low counting probability). The cause of the three sum peaks is discussed in detail in Section 3.4.

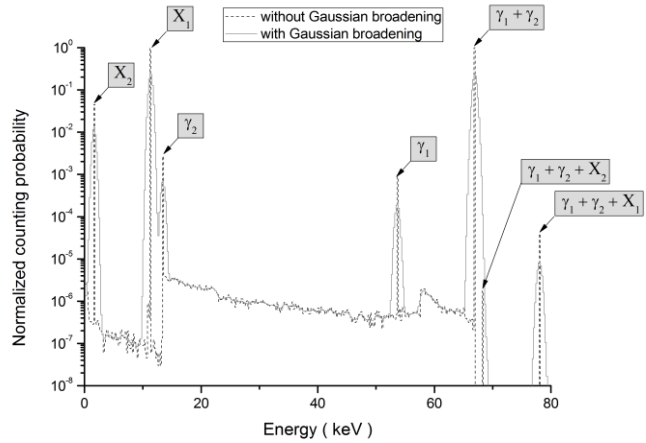

**Fig. 7.**  Simulated spectrum of $^{73}$As

Fig. 8 is the simulated background spectrum of internal $^{71}$Ge. Except for three X-ray peaks, no sum peaks exist in the spectrum, because no β-rays or γ-rays are emitted in the EC decay of $^{71}$Ge. The X-rays are impossible to sum-up, because only one X-ray is emitted in each EC decay.

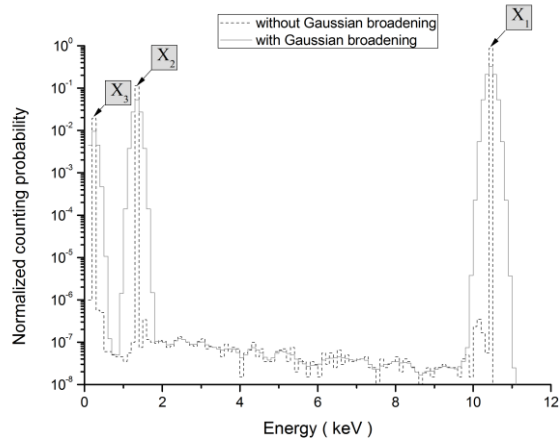

**Fig. 8.**  Simulated spectrum of $^{71}$Ge

## 3.4 The impact of detector response time

Taking $^{73}$As as an example. As Fig. 9 shows, the level 2 state of $^{73}$As has a quite long half-life, 0.499 s, so the X-rays and gamma do not necessarily sum-up.

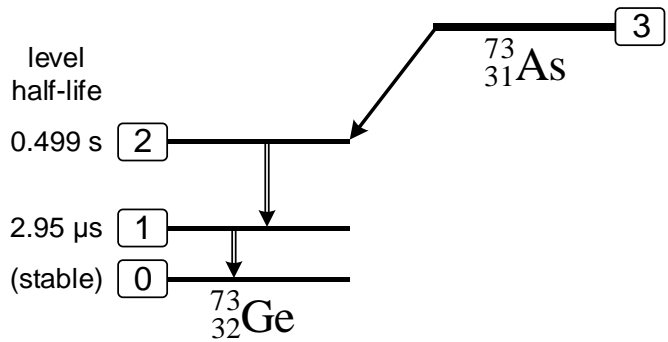

**Fig. 9.** Decay scheme of $^{73}$As, showing the half-lives for each level

Fig. 10 shows that the peaks in the spectrum change with detector response time. Fig. 10(a) shows there is no coincidence summing-up, because when the response time = 0 (ideal situation), the detector can separate all the individual particles. Fig. 10(b) shows that if the response time becomes longer, the $X_1$ and $X_2$ peaks do not change, but the $\gamma_1$ and $\gamma_2$ peaks partially sum-up. Fig. 10(c) shows that if the response time = 2.95 μs, exactly 50% of $\gamma_1$ and $\gamma_2$ sum-up. Fig. 10(d) shows nearly all of $\gamma_1$ and $\gamma_2$ summing-up and forming a sum peak. Similarly, as the response time becomes even longer, as Fig. 10(e)~(g) show, the X-rays also begin to sum-up. Fig. 10(h) shows each of the two X-rays totally summing-up with the gamma sum peak, when the response time = ∞ (ideal situation). The simulated $^{73}$As spectrum therefore needs to be checked against the experimental spectrum, which is useful for obtaining the true response time of the germanium detector.

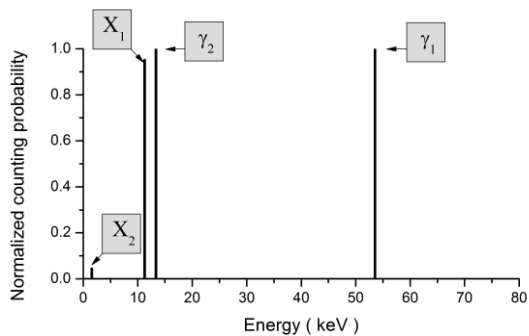

(a) Response time = 0

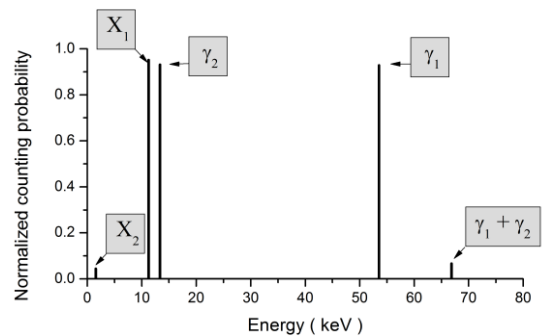

(b) Response time = 0.3 μs

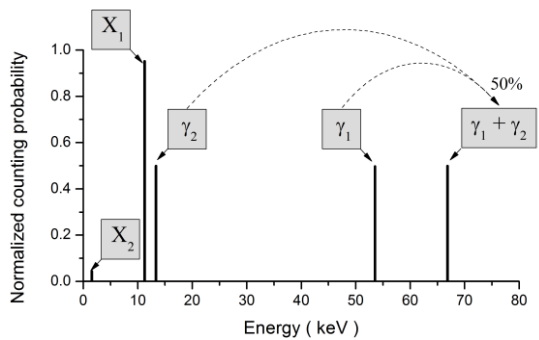

(c) Response time = 2.95 μs

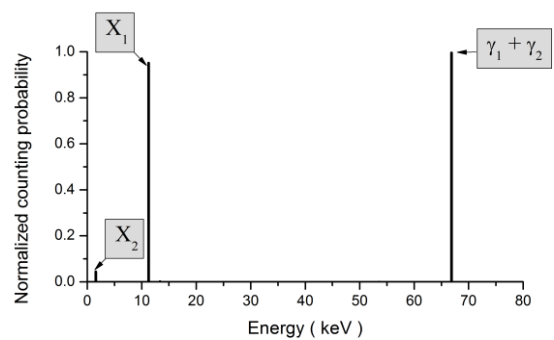

(d) Response time = 30 μs

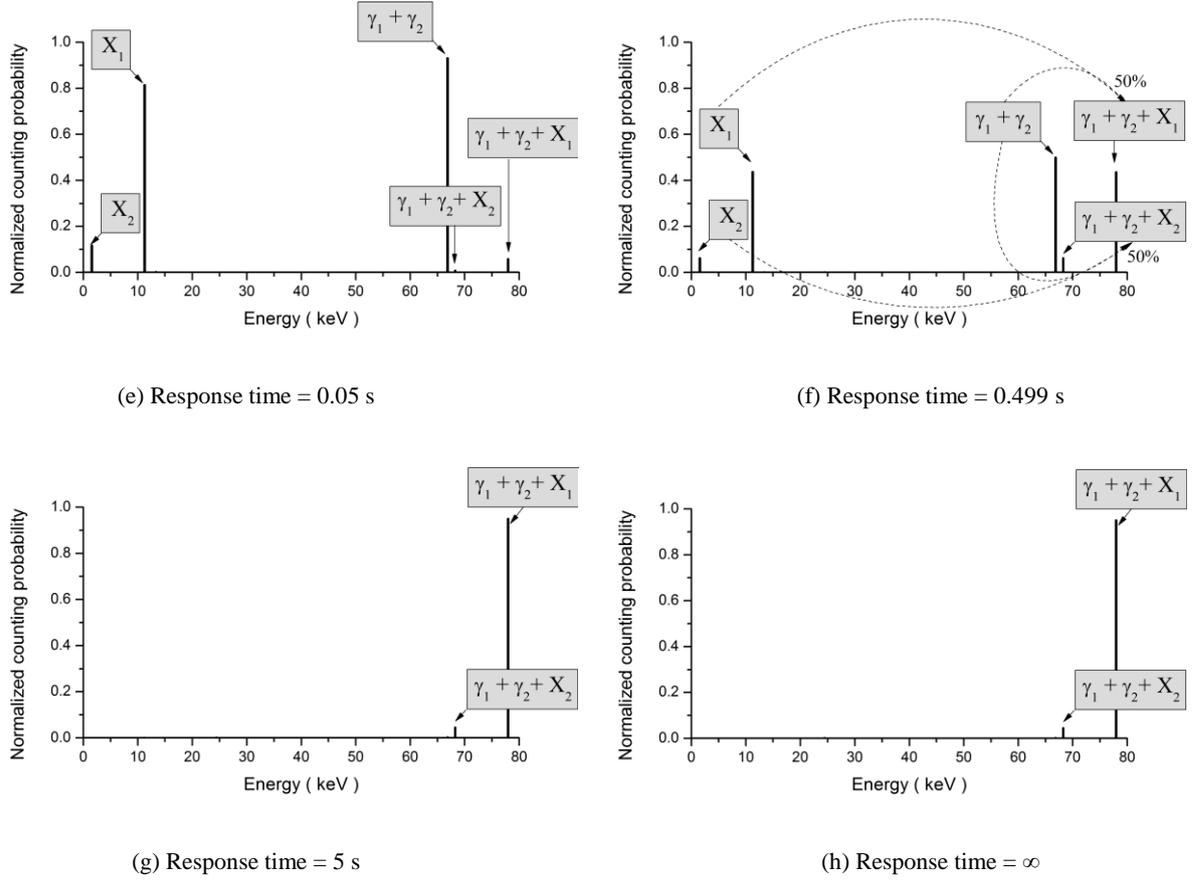

(e) Response time = 0.05 s  (f) Response time = 0.499 s

(g) Response time = 5 s  (h) Response time = ∞

**Fig. 10.**  Simulated spectrum of $^{73}$As with various detector response time

# 4  Conclusion

Cosmogenic nuclides exist inside germanium crystals and contribute to background noise in germanium-based detectors; this is an important background source for dark matter experiments and neutrinoless double-beta decay research. In extra-low background experiments, the internal cosmogenic nuclide background is a problem which cannot be ignored, as long as the measured particle and target material are in the same crystal.

Due to the possibility of evaluating the background contribution from these cosmogenic nuclides in the high energy region, based on their X-ray spectra in the low energy region, we propose a simulation method to deal with the coincidence summing-up effect, based on our use of graph theory to model the decay and level transitions model. The benefits of this method can be summarized as:

• Sampling the transition time to decide summing-up, instead of using a coincidence factor to evaluate the coincidence summing-up effect;

• the whole energy region spectrum (from 0 to 4566 keV, which is the $Q$ value of $^{56}$Co) is evaluated;

• the detector response time is taken into consideration and can be changed according to the actual situation.

By using this and referring to the measured spectrum, the cosmogenic nuclide background can be quantitatively evaluated within the range 0 ~ 4566 keV.

Though we established this method to deal with the coincidence summing-up effect, there are still several uncertain parameters, such as the detector response time. In our simulations, we assumed a response time of 30 μs, but the true response time is decided by several complex factors, including electric field distribution, electronics and so on. In future, further research will be conducted on the response time, which should result in more accurate simulated spectra.

## Acknowledgments

This work is supported by the National Natural Science Foundation of China (No.11175099, No.10935005, No.10945002, No.11275107) and National Basic Research Program of China (973 Program) (2010CB833006).